\documentclass[%
preprint,
nofootinbib,
 amsmath,amssymb,
 aps,
]{revtex4}

\usepackage{multirow}
\usepackage{booktabs}
\usepackage{cancel}
\usepackage{color}
\usepackage{slashed}
\usepackage{graphicx}
\usepackage{float}
\usepackage{dcolumn}
\usepackage{bm}
\usepackage{subfigure}

\begin{document}

\title{Monojet search for heavy neutrinos at future $Z$-factories}

\author{Yin-Fa Shen$^{1}$\footnote{syf70280@hust.edu.cn}, Jian-Nan Ding$^{2}$\footnote{dingjn13@lzu.edu.cn, corresponding author}, Qin Qin$^{1}$\footnote{qqin@hust.edu.cn, corresponding author}}

\address{
$^1$School of Physics, Huazhong University of Science and Technology, Wuhan 430074, China \\
$^2$School of Nuclear Science and Technology,  Lanzhou University, Lanzhou 730000,  China}

\begin{abstract}

The existence of heavy neutrinos is the most prominent feature of type-I seesaw-like mechanisms, which naturally explain tiny masses of observed neutrinos. 
Various experiments have been conducted to search for heavy neutrinos, with meson decays sensitive to heavy neutrinos lighter than $B$ mesons and high energy collider experiments sensitive to the ones with masses larger than $\mathcal{O}(0.1)$ TeV. 
In between, there is a mass gap uncovered by these two kinds of experiments, and by this work, we investigate the 3-15 GeV neutrino searches at future $Z$-factories via the process $e^{-}e^{+}\rightarrow Z \rightarrow \nu N \rightarrow \nu \ell\bar{q}q'$ with $\ell=e,\mu$. 
Based on the simulation of signal and background events, we find that the signals appear as monojet events owning to the large Lorentz boost of the heavy neutrinos. Furthermore, the substructures of such monojet are essential to reconstruct the signal and suppress the background.  
The upper bounds on the cross section of the signal process and the mixing parameter ${\left| V_{\ell N} \right|}^2$ are obtained. 
Compared with other experiments, the monojet method at future $Z$-factories will be able to fill the gap around this mass range.
\end{abstract}

\maketitle

\section{Introduction}
Searching for new physics beyond the Standard Model (SM) is a crucial task in particle physics. Actually, there have been clear hints beyond the minimal SM in the neutrino sector, {\it i.e.}, the neutrino oscillations~\cite{Super-Kamiokande:1998kpq,SNO:2002tuh,DayaBay:2012fng}, which indicate nonzero but tiny neutrino masses~\cite{Mohapatra:2006gs}. Although the neutrino masses can be induced by the Yukawa mechanism analogous to the quark sector, the tiny Yukawa couplings will be quite unnatural~\cite{Bilenky:2014uka}. There have been various theories trying to naturally explain the tiny neutrino masses~\cite{Minkowski:1977sc,Mohapatra:1979ia,Yanagida:1980xy,Schechter:1980gr,delAguila:2008cj,Atre:2009rg,Magg:1980ut,Cheng:1980qt,Lazarides:1980nt,Mohapatra:1980yp,Foot:1988aq,Altarelli:2002hx,Arkani-Hamed:1998wuz,Drees:1997id,CentellesChulia:2020dfh}. A typical idea is that if heavy Majorana neutrinos are introduced, the light neutrinos will obtain masses inversely proportional to the large Majorana mass scale. This is the so-called seesaw mechanism~\cite{Minkowski:1977sc,Mohapatra:1979ia,Yanagida:1980xy,Schechter:1980gr,CentellesChulia:2020dfh,delAguila:2008cj,Atre:2009rg}. 
Its most prominent phenomenological feature is the existence of heavy neutrinos. Therefore, searching for such heavy neutrinos is crucial for verifying the seesaw mechanism and probing the origin of neutrino masses.

In different seesaw-inspired models, heavy neutrino masses can span from eV to $10^{16}$ GeV \cite{Drewes:2015jna}. Among these different mass ranges, heavy neutrinos with $\mathcal{O}(10)$ GeV masses have some specific interesting features. For example, 
heavy neutrinos in the mass range of 1-12 GeV can provide a proper answer simultaneously to the puzzle of neutrino masses, dark matter, and the baryon asymmetry~\cite{Asaka:2005pn}. Practically, such a mass range can be explored directly by proposed $Z$-factories~\cite{CEPCStudyGroup:2018ghi,Baer:2013cma,Gomez-Ceballos:2013zzn} in the near future or even the facilities like the Large Hadron Collider (LHC)~\cite{Cvetic:2019shl,CMS:2018iaf,CMS:2022fut,ATLAS:2019kpx} at present. 

Various experimental searches for heavy neutrinos have been carried out, but no signals have been observed yet and thus only constraints on theoretical parameters of the seesaw mechanism are imposed~\cite{T2K:2019jwa,BESIII:2015tql,NA62:2017qcd,PIENU:2019usb,PIENU:2017wbj,Belle:2013ytx,NA62:2017ynf,E949:2014gsn,Castro:2013jsn,Wang:2014lda,LHCb:2014osd,CHARM:1985nku,CHARMII:1994jjr,WA66:1985mfx,Yuan:2013yba,BESIII:2019oef,Bryman:2019bjg,GERDA:2020xhi,Deppisch:2020ztt,ALEPH:1991qhf,DELPHI:1996qcc,CMS:2018iaf,CMS:2022fut,ATLAS:2019kpx,Bolton:2019pcu}. The experiments can be roughly categorized into direct and indirect searches. Indirect searches include lepton number violating (LNV) decays of mesons~\cite{Yuan:2013yba,BESIII:2019oef} and neutrinoless double beta  $(0\nu\beta\beta)$ decays in certain nuclei~\cite{GERDA:2020xhi,Deppisch:2020ztt}. 
Also, precision SM tests like the $Z$ boson decay width can be used to set constraints indirectly~ \cite{Abada:2014cca,Das:2018hph,Abada:2015zea}.
As for direct searches, different experimental methods are efficient for heavy neutrinos in different mass ranges. The most stringent constraints on heavy neutrinos below 5 GeV are given by searches in meson decays~\cite{Bryman:2019bjg,T2K:2019jwa,BESIII:2015tql,NA62:2017qcd,PIENU:2019usb,PIENU:2017wbj,Belle:2013ytx,NA62:2017ynf,E949:2014gsn,Castro:2013jsn,Wang:2014lda,LHCb:2014osd,CHARM:1985nku,CHARMII:1994jjr,WA66:1985mfx}. 
Heavier neutrinos are typically hunted through decays of the $Z$, $W$, and $H$ bosons or direct productions at high energy colliders~\cite{Keung:1983uu,Petcov:1984nf}. Among these experiments, the LHC~\cite{CMS:2018iaf,ATLAS:2019kpx,Cvetic:2019shl} is collecting more and more precise data for the $\mathcal{O}$(100) GeV mass range. Although a recent CMS search for $\mathcal{O}$(10) GeV heavy neutrinos has set a new stringent constraint on the mixing parameters~\cite{CMS:2022fut}, it mostly focus on long-lived heavy neutrinos, which is more model-dependent as we will discuss later. Hence, we will concentrate on the capacity of future $Z$-factories to search for heavy neutrinos within the 3-15 GeV mass range. Studies for heavier neutrino searches at $Z$-factories can be found in {\it e.g.}~\cite{Blondel:2014bra,Antusch:2016ejd,Antusch:2016vyf,Antusch:2017pkq,Ding:2019tqq,Gao:2021one,Blondel:2021mss,Abada:2014cca,Banerjee:2015gca,Liao:2017jiz,Blondel:2018mad,Graverini:2016mhd,Antusch:2015mia,Alekhin:2015byh,Graverini:2015dka,Blondel:2021ema,Knapen:2021svn,Wang:2019xvx}.

At $Z$-factories, 3-15 GeV neutrinos lighter than the $Z$ boson are mainly produced through the process $e^{-}e^{+}\rightarrow Z \rightarrow \nu N$, and we consider their reconstruction via the decay channel $N \rightarrow \ell W^{*} \rightarrow \ell q'\bar{q}$, in which all the decay products are visible. As the heavy neutrinos are highly boosted, their decay products basically stay collinear to each other. Therefore, the signal events are recognized to contain one jet and missing energy, {\it i.e.}, the monojet events. It turned out that the jet substructure provided essential information in distinguishing the signal from the background. Assuming three different integrated luminosities (0.1, 1 \emph{and} 10 $\rm ab^{-1}$) of future $Z$-factories, we found that the capacity of finding such heavy neutrinos could be improved by 
one to three orders of magnitude compared to the DELPHI~\cite{DELPHI:1996qcc}, especially for the 4-10 GeV mass range. Comparisons are also made with other experimental results.

The paper will be organized as follows. In Section II, the general formulation of the seesaw mechanism will be introduced, and the relevant signal process will be briefly discussed. In Section III, we will illustrate the simulation of the signal and background events and the signal selection conditions to distinguish the signal from the background. In Section IV, the results will be presented and discussed. We will conclude with Section V.

\section{Theoretical set up}
Following the spirit of the seesaw mechanism and gauge invariance, the simplest renormalizable Lagrangian terms involving neutrino masses read~\cite{delAguila:2008cj,Atre:2009rg}:
\begin{equation}
\mathcal{L}_{\nu} \ni \frac{1}{2}i\bar R^{c}_{i}\slashed{\partial}R_{i}-y_{ij}\bar R_{i} \tilde \Phi L_{j}-\frac{1}{2}(M_{N})_{ij} \bar R^{C}_{i}R_{j}+h.c.,
\end{equation}
where $\tilde \Phi=i\sigma_{2}\phi^{*}$ with $\phi$ being the Higgs doublet and $\sigma_{2}$ being the second Pauli matrix, $L_j$ is an SU(2) doublet of leptons, $y$ is the leptonic Yukawa coupling matrix, $M_{N}$ is the Majorana mass matrix, $C$ means the charge conjugation, and $R_{i}$ $(i=1,...,n)$ are certain SU(2) singlet right-handed neutrino fields. The number of heavy right-handed neutrinos depends on whether the lightest neutrino is massless or not~\cite{Wyler:1982dd}. The Majorana mass terms are generated by new dynamics of higher scales. 
The right-handed neutrino fields $R_{j}$ are SU(2) singlet, and their hypercharges must be zero for the nature of Majorana fermions. Such conditions will maintain the cancellations of anomalies in SM and predict the fermions' correct electric charges (see~\cite{Babu:1989tq} for more details).

After the spontaneous electroweak symmetry breaking, the Dirac mass $(M_{D})_{ij}$ of the neutrinos are generated from the Yukawa couplings, and the total mass matrix reads
\begin{equation}
\begin{bmatrix}
0 & M_{D}\\
M^{T}_{D} & M_{N}
\end{bmatrix},
\end{equation}
with respect to the neutrino flavor eigenstates $\{\nu_{L},R\}$. Diagonalizing the mass matrix, we will get $3$ light ($\nu_i$) and $n$ heavy ($N_j$) mass eigenstates, whose interaction terms are given by
\begin{equation}\label{eq:l}
\begin{split}
\mathcal{L} \ni \sum_{\ell,i,j}-\frac{g}{\sqrt{2}}(U^{*}_{\ell i}\bar \nu_{i}+ V^{*}_{\ell j}\bar N_{j}) W^{+}_{\mu} \gamma^{\mu}P_L\ell+h.c. \qquad\ \ \ \\
-\frac{g}{2\cos\theta _{W}}(U^{*}_{\ell i}\bar \nu_{i}+ V^{*}_{\ell j}\bar N_{j}) Z_{\mu} \gamma^{\mu}P_L(U_{\ell i}\nu_{i}+V_{\ell j}N_{j}),
\end{split}
\end{equation}
where $U_{\ell i}$'s ($V_{\ell j}$'s) are the mixing parameters between the weak interaction eigenstates and the light (heavy) mass eigenstates,  $\theta_W$ is the weak mixing angle, and $P_L$ is the left-handed projector. Practically, the mixing parameter $V_{\ell j}$'s are always so tiny \cite{Minkowski:1977sc,Schechter:1980gr} that the $U_{\ell i}$ matrix is quasi-unitary, {\it i.e}. $U^\dagger U\approx \mathbf{1}_{3\times3}$.

Let us consider the production process $e^{-}e^{+}\rightarrow Z\rightarrow \nu_{\ell} N_{j}$ and the decay process $N_{j}\rightarrow \ell W^{*}$ of a heavy neutrino at $Z$-factories, as shown in FIG.~\ref{Fig2}. If we sum over contributions from all light neutrinos in the production, the cross section can be formulated as $\sigma(e^{-}e^{+}\rightarrow Z\rightarrow \nu_{\ell}N_{j}) \propto \sum_{\ell} \sum_{i=1}^{3}\left|U^{*}_{\ell i}V_{\ell j}\right|^{2}=\sum_{\ell} \sum_{i=1}^{3}V_{\ell j}^{*}U_{\ell i}U^{*}_{\ell i}V_{\ell j}\approx \sum_{\ell} \left|V_{\ell j}\right|^{2}$. On the other hand, as shown in \cite{Atre:2009rg} , the total decay width of a heavy neutrino lighter than the $W$ boson can be expressed as $\Gamma_N=\sum_{\ell}C_\ell \left|V_{\ell N}\right|^2$ where the coefficients $C_\ell$ can be approximately regarded to be universal for different flavors if $M_N^2\gg m_\ell^2$ is provided. It leads to the total decay width of a heavy neutrino $\Gamma_N\propto \sum_\ell \left|V_{\ell N}\right|^2$ and the branching ratio $BR(N_{j}\rightarrow \ell W^{*})\propto \frac{\left|V_{\ell j}\right|^{2}}{\sum_{\ell} \left|V_{\ell j}\right|^{2}}$. Thus we can approximately conclude that the total cross section of the process $\sigma(e^{-}e^{+}\rightarrow Z \rightarrow \nu_{\ell}N_{j}\rightarrow \nu_{\ell}\ell W^{*}) \propto \left|V_{\ell j}\right|^{2}$, indicating it is sensitive to $\left|V_{\ell j}\right|^{2}$ and almost independent on other mixing parameters.

\section{Simulation and event selection}
\begin{figure}
\centering
\includegraphics{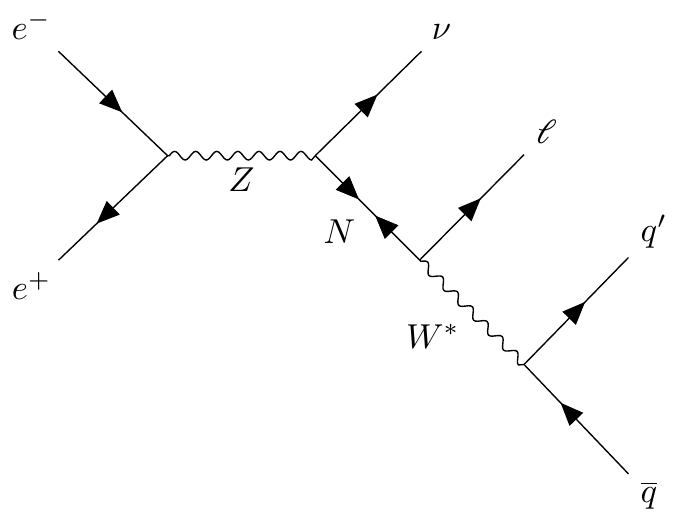}
\caption{The Feynman diagram of the signal process, with $\ell=e,\mu$ and $\nu$ summing over three active neutrino flavors.}
\label{Fig2}
\end{figure}

\begin{figure}
\centering
\includegraphics{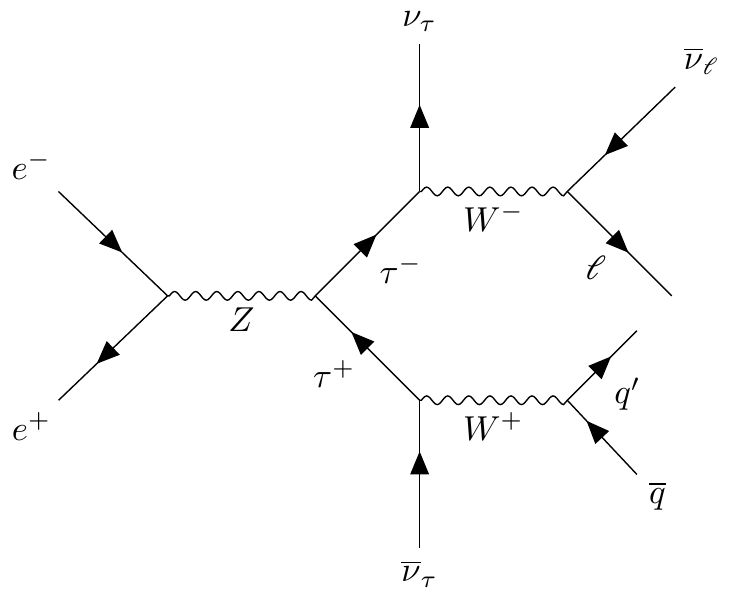}
\caption{The Feynman diagram of the $\tau^+\tau^-$ background.}
\label{background}
\end{figure}

In this section, we will perform the simulation for the signal and background events and find the proper event selection conditions so that the signal events can be reconstructed out of the SM background efficiently. For the signal, the heavy neutrinos $N_{j}$ are mainly produced via the process $e^{-}e^{+}\rightarrow Z \rightarrow \nu_{\ell} N_{j}$ instead of $e^{-}e^{+}\rightarrow Z \rightarrow N_{i}N_{j}$, because the latter is more suppressed by $|V_{\ell j}|^{2}$ compared to the former. We choose the semi-leptonic decay channels $N_{j} \rightarrow \ell \bar{q}q'$ to reconstruct $N_{j}$, whose decay products can be fully detected in principle, as displayed in FIG.\ref{Fig2}.
As only one heavy neutrino will be involved in this study, we use $M_{N}$ and $V_{\ell N}$ referring to $M_{Nj}$ and $V_{\ell j}$ from now on. For the background, all the two-fermion and four-fermion processes are considered, and it turns out that the dominant process is $e^{-}e^{+}\rightarrow Z \rightarrow \tau^+\tau^-$ for its overwhelming cross section and signal-like decay modes. As shown by FIG.~\ref{background}, the $\tau^+\tau^-$ pair can decay into one charged lepton, quarks, and missing energy in the final state, just as the signal process. 
The signal and background events are simulated and analyzed using FeynRules~\cite{Alloul:2013bka} with the HeavyN package~\cite{Alva:2014gxa,Degrande:2016aje}, MadGraph5\_aMC@NLO~\cite{Alwall:2014hca,Frederix:2018nkq}, MadWidth~\cite{Alwall:2014bza},  Pythia8~\cite{Sjostrand:2014zea}, and Delphes3 with FastJet~\cite{deFavereau:2013fsa,Cacciari:2011ma}. More specifically, we adopt the \emph{anti-$k_T$} algorithm with the angular distance $\Delta R <0.5$ to reconstruct the \emph{inclusive} jet.

As mentioned previously, the large Lorentz boost of a heavy neutrino produced at the $Z$ mass pole implies that all its decay products, including a charged lepton and hadrons, will fly in a collinear direction and result in a monojet signature. Therefore, the signal events should consist of a monojet with a charged lepton inside and large missing energy in the opposite direction.

Besides the monojet structure, the signal events also have specific kinematic features to distinguish them from the background further. 
Apparently, the opposite side of the monojet will be empty because there is only missing energy.
The monojet and the missing energy fly back-to-back and almost divide the total collision energy equally due to the small heavy neutrino mass. 
Meanwhile, the reconstructed center-of-mass energy $E_\text{cm-rec}=\slashed E + E_{j}$ will locate around $E_\text{cm}=91.2\,{\rm GeV}$, where the missing energy $\slashed E$ is supposed to be the norm of the missing momentum solved by total  conservation. As for the background, such as $\tau^+\tau^-$, there are typically more than one non-collinear neutrinos, and hence the reconstructed center-of-mass energy will be shifted. 
Moreover, the substructure of the monojet can provide more information to suppress the background. First, the signal monojet typically contains one hard charged lepton. 
Also, the momentum distributions in the monojets of the signal are different from those of the background. Taking the $\tau^+\tau^-$ background as an example--the $\tau$ lepton is even lighter than $N$, so the boost is more drastic, and the shape of the monojet in the final state will be more collimating.

\begin{table}
\caption{
The event selection conditions and their cumulative “ bkg (signal) efficiencies”, which means the percentage of the background (signal) events surviving after a particular selection condition and the ones above it are all applied. We take the $5\,\mathrm{GeV}$ heavy neutrino with $\ell =e$ as the example.
}\label{table:cuts}
\resizebox{\textwidth}{!}{
\begin{tabular}{cccc}
\toprule
\hline
\multirow{2}{*}{\textbf{ }} & \multirow{2}{*}{\textbf{selection condition}} & \textbf{bkg efficiency} & \textbf{signal efficiency} \\ 
 & &\textbf{(cumulative)}& \textbf{(cumulative)} \\
 \hline
event type & monojet & 15.44695\% & 78.45346\%\\
\hline
\multirow{2}{*}{missing energy} & $ \slashed E_{T} > 23\,\rm GeV $ & 0.62255\% & 75.30132\% \\
                     & $ 40\ {\rm GeV} < \slashed E < 50\,\rm GeV $ & 0.03372\% & 68.40370\% \\
                     \hline
\multirow{4}{*}{jet} & $ p^{j}_{T} > 23\,\rm GeV $ & 0.03337\% & 67.81746\% \\
                     & $ 30\, {\rm GeV} < E_{j} < 50\, \rm GeV $ &  0.02902\% & 67.10762\% \\
                     & TauTagging = 0  &  0.02423\% & 67.02997\% \\
                     & $ 3 < \Delta R_{\slashed {E} j} < 4.1  $ & 0.02194\% & 66.82435\% \\
                     \hline
\multirow{3}{*}{substructure} & $ \ell =e\ \emph{and}\ \Delta R_{j\ell} < 0.5 \ \emph{and} \ p^{\ell}_{T} > 5 \, \rm{GeV} $ & 0.00053\% & 52.20332\%  \\
                             & $ \Delta R_j < 0.11:\frac{\sum p_{T} + \sum E_{T}}{p^{j}_{T}} < 90\% $ & 0.00050\% & 48.54479\% \\
                             & $ \Delta R_{\slashed E} < 2:\sum p_{T} + \sum E_{T} < 0.3 \, {\rm GeV} $ & 0.00046\% & 48.50213\% \\ 
                             \hline
others & $ \left| E_\text{cm-rec} - E_\text{cm} \right| < 20 \, {\rm GeV} $ & 0.00046\% & 48.50213\% \\
\hline
\bottomrule
\end{tabular}
}
\end{table}

\begin{figure}
\centering
\subfigure{
\centering
\includegraphics[width=6cm]{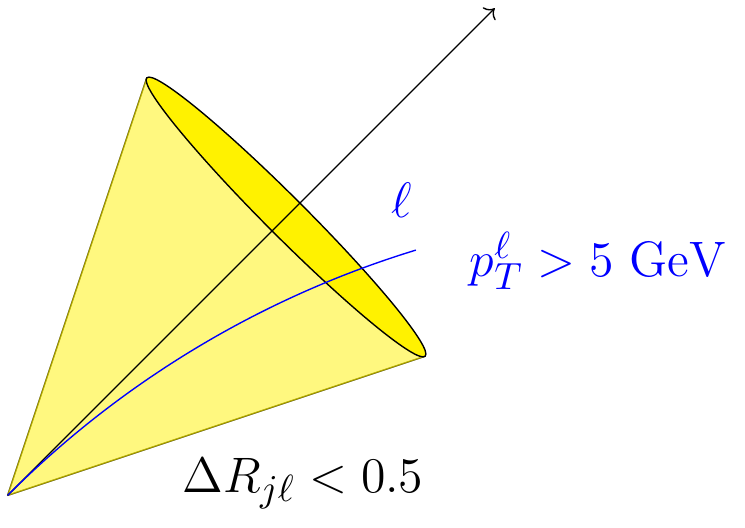}
}
\subfigure{
\centering
\includegraphics[width=7.4cm]{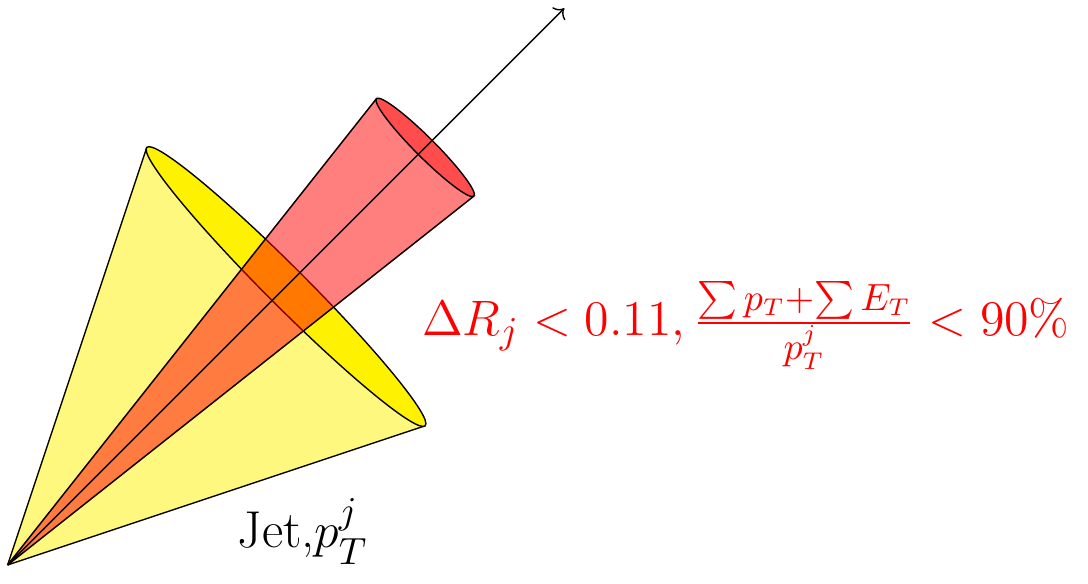}
}
\caption{Sketches to visualize the event selection conditions about the jet substructure listed in TABLE~\ref{table:cuts}. The yellow cone stands for a jet in the arrow direction. The left panel shows that the hard lepton contained in the jet should satisfy the two conditions $\Delta R_{j\ell}<0.5$ and $p_T^\ell>5$ GeV. In the right panel, the red cone includes the particles with angular separations smaller than 0.11 from the jet direction, and they in total are required to account for less than 90\% of the transverse jet momentum.}
\label{substructure}
\end{figure}

Based on the above analysis and tests to the simulated events, we suggest using the event selection conditions listed in TABLE~\ref{table:cuts}.
First, only the monojet type of events is selected, containing only one jet $(j)$ without isolated charged leptons. The conditions of the ``missing energy'' category and the first two conditions of the ``jet'' category guarantee that the events are hard processes with large transverse momenta and energies, where $\slashed{E}$ and $\slashed{E}_T$ are the missing energy and its transverse component while $E_j$ and $p^j_T$ are the energy and transverse momentum of the jet. We also use the condition ``TauTagging=0'', which claims that no jet is tagged as a $\tau$ jet to suppress the $\tau^+\tau^-$ background further.
Because the monojet and the missing energy fly back-to-back, the azimuthal angle between them ($\Delta \phi_{\slashed E j}$) is about $\pi$, and the angular separation $\Delta R_{\slashed E j}=\sqrt{\Delta \phi^2_{\slashed E j}+\Delta \eta^2_{\slashed E j}}$ must have the corresponding threshold, where $\Delta \eta^2_{\slashed E j}$ is the square of the pseudorapidity difference between them. 
The first condition of ``substructure'' is actually the restriction on the charged lepton $\ell$ inside the monojet, with $\Delta R_{j\ell}$ being the angular separation between the jet direction and the charged lepton and $p^{\ell}_T$ being the transverse momentum of the charged lepton. 
It should be noted that the events containing only one electron (or muon) satisfying the $\Delta R_{j\ell}$ and $p^{\ell}_T$ conditions at the same time are adopted. The charged lepton is required to be hard to avoid the pollution from leptonic decays of quarks. 
The other two conditions using the transverse momenta of charged particles $p_T$ and the transverse energies of neutral particles $E_T$ are set based on the momentum distribution of the particles in the jet and around the missing energy direction, as mentioned in the last paragraph. 
The first two conditions of ``substructure'' are visualized in FIG.~\ref{substructure}. 
At last, the ``other'' part is the restriction on the reconstructed center-of-mass energy $E_\text{cm-rec}$. For monojet events it should locate around the collision energy $E_\text{cm}=91.2\,\mathrm{GeV}$. As shown in TABLE~\ref{table:cuts}, after all these selection conditions are applied, 99.99954\% background events are excluded, while almost half of the signal events still survive for the $M_N=5$ GeV case.


\section{Result}

Applying the selection conditions to the simulated signal and background events, we can investigate the ability of future $Z$-factories to hunt the ``light" heavy neutrinos with masses in the range of 3-15 GeV. Analogous to~\cite{Ding:2019tqq}, we consider that the significance $s$ defined by
\begin{equation}
s\equiv \frac{N_{s}}{\sqrt{N_{B}+N_{s}}},\label{s1}
\end{equation}
is set to about 1.7 to obtain the corresponding upper bounds at 95\% confidence level (CL), where $N_s$ and $N_B$ are the numbers of signal and background events satisfying the selection conditions, respectively. 

\begin{figure}
\centering
\subfigure[The electron case;]{
\centering
\includegraphics[width=7.9cm]{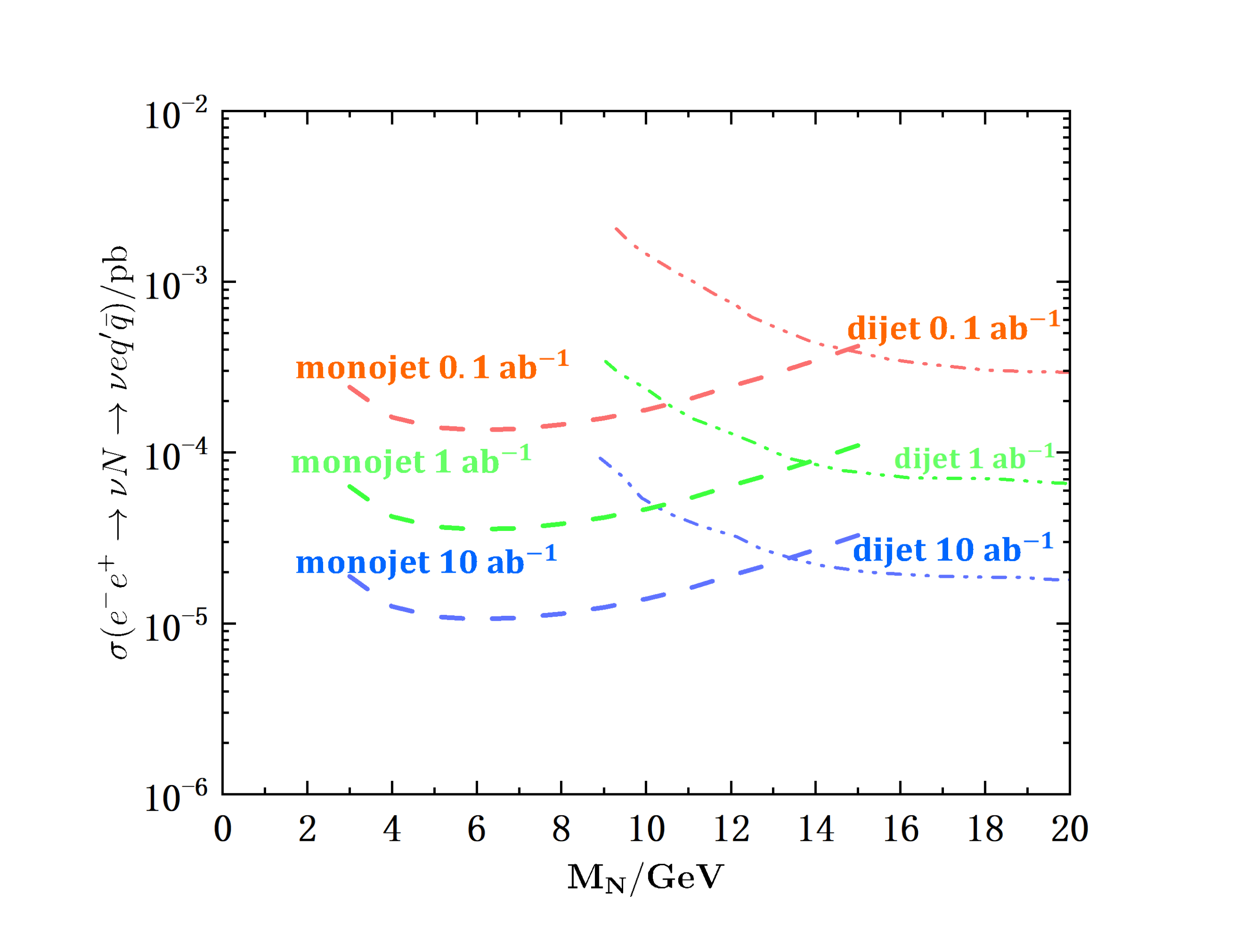}
\label{sigmae}
}
\subfigure[The muon case.]{
\centering
\includegraphics[width=7.9cm]{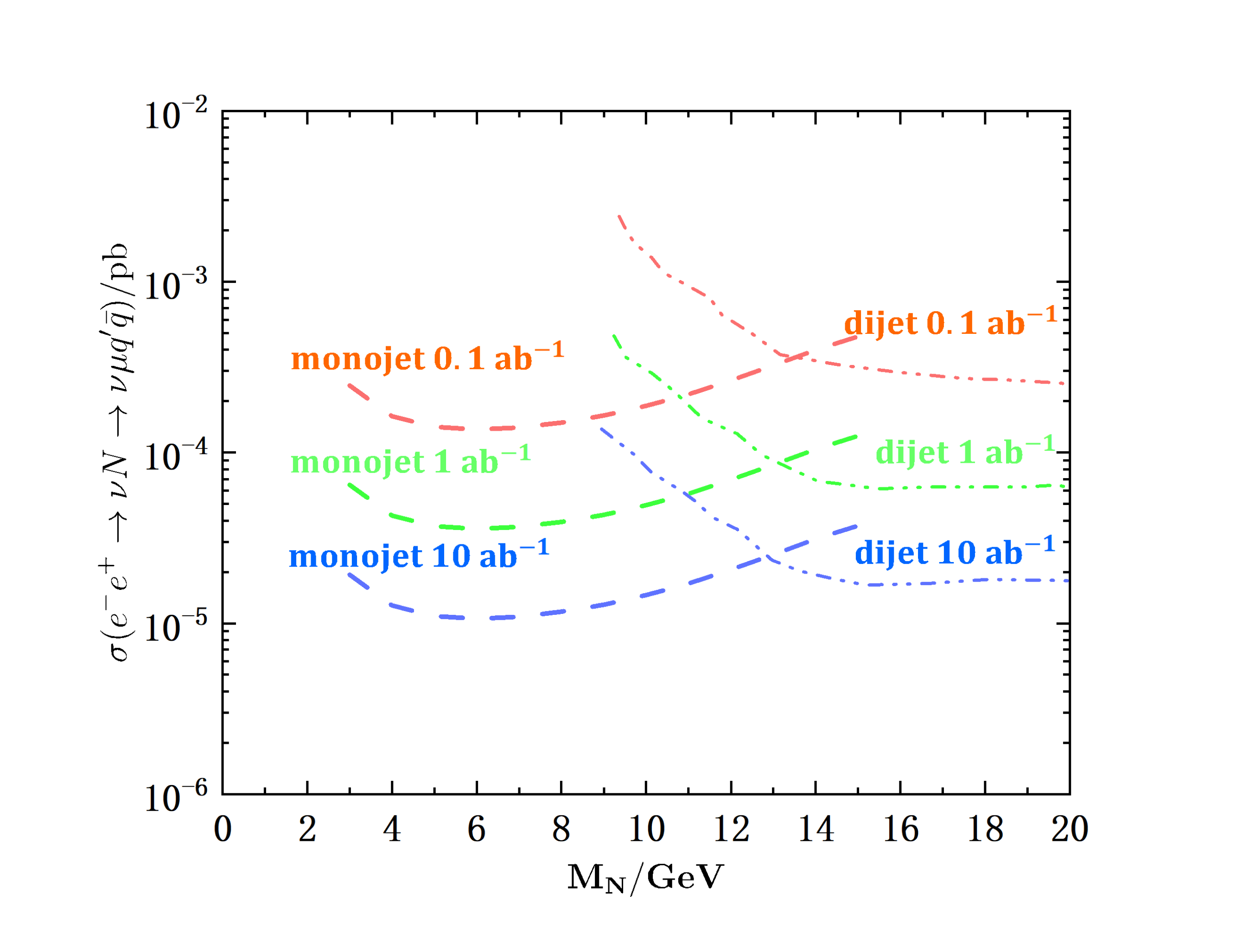}
\label{sigmamu}
}
\caption{The upper bounds on the cross sections $\sigma(e^+e^-\to \nu N\to \nu \ell q'\bar{q})$ at 95\% CL given by future $Z$-factories with three setups of integrated luminosities $L=0.1,1\ \emph{and}\ 10\ \text{ab}^{-1}$, using the monojet and dijet~\cite{Ding:2019tqq} methods.}
\label{sigma}
\end{figure}

The upper bounds on the cross sections $\sigma(e^+e^-\to \nu N\to \nu \ell q'\bar{q})$ with 3 GeV $<M_N<$ 15 GeV and $\ell=e,\mu$ obtained in our analysis are displayed in FIG.~\ref{sigma}, with three setups of integrated luminosities $L=0.1,1\ \emph{and}\ 10\ \text{ab}^{-1}$. The corresponding upper bounds for the 10-20 GeV heavy neutrinos obtained in \cite{Ding:2019tqq} are also shown as a comparison, where final states containing two jets and a lepton are used to reconstruct the heavy neutrinos. It can be seen that the monojet method adopted in our analysis has a good performance around the 5-10 GeV mass range. For lighter neutrinos, the shape of a monojet will be more collimating as a $\tau$-jet, so the second selection condition of ``substructure" will lower the signal reconstruction efficiency and make the constraints looser. For heavier neutrinos, the monojet method gradually loses its efficacy because of the reduction of the Lorentz boost, and the dijet method has the advantage. It is observed that the monojet bounds for $M_N\approx$ 5 GeV and the dijet bounds for $M_N>$ 15 GeV are comparable to each other. This behavior is consistent with that observed in the DELPHI search~\cite{DELPHI:1996qcc}. For the mass range 9 GeV $<M_N<$ 14 GeV, where the two methods both work less well, they can be considered to be combined for setting a more stringent constraint.

\begin{figure}
\centering
\subfigure[The electron case;]{
\centering
\includegraphics[width=7.9cm]{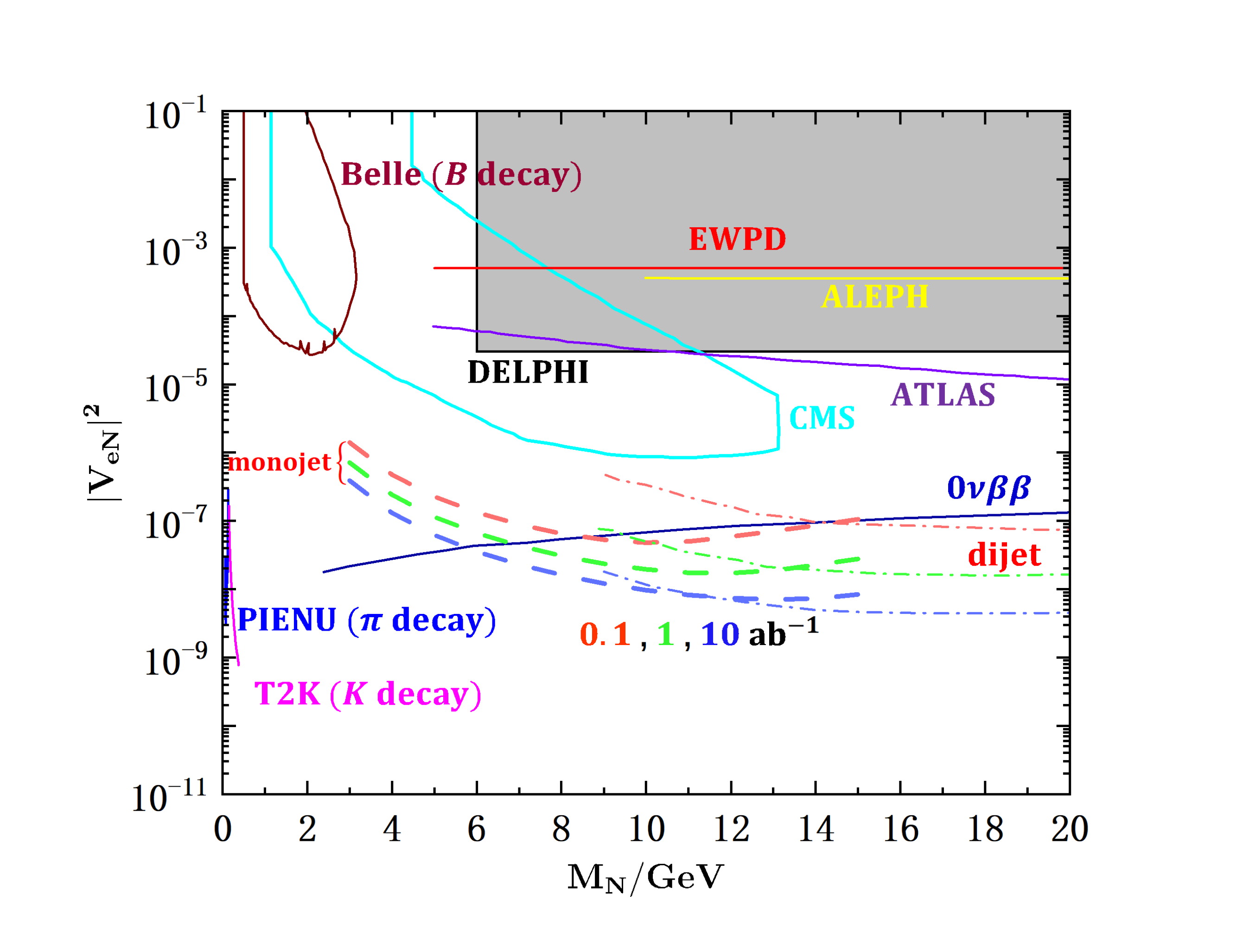}
\label{VeN}
}
\subfigure[The muon case.]{
\centering
\includegraphics[width=7.9cm]{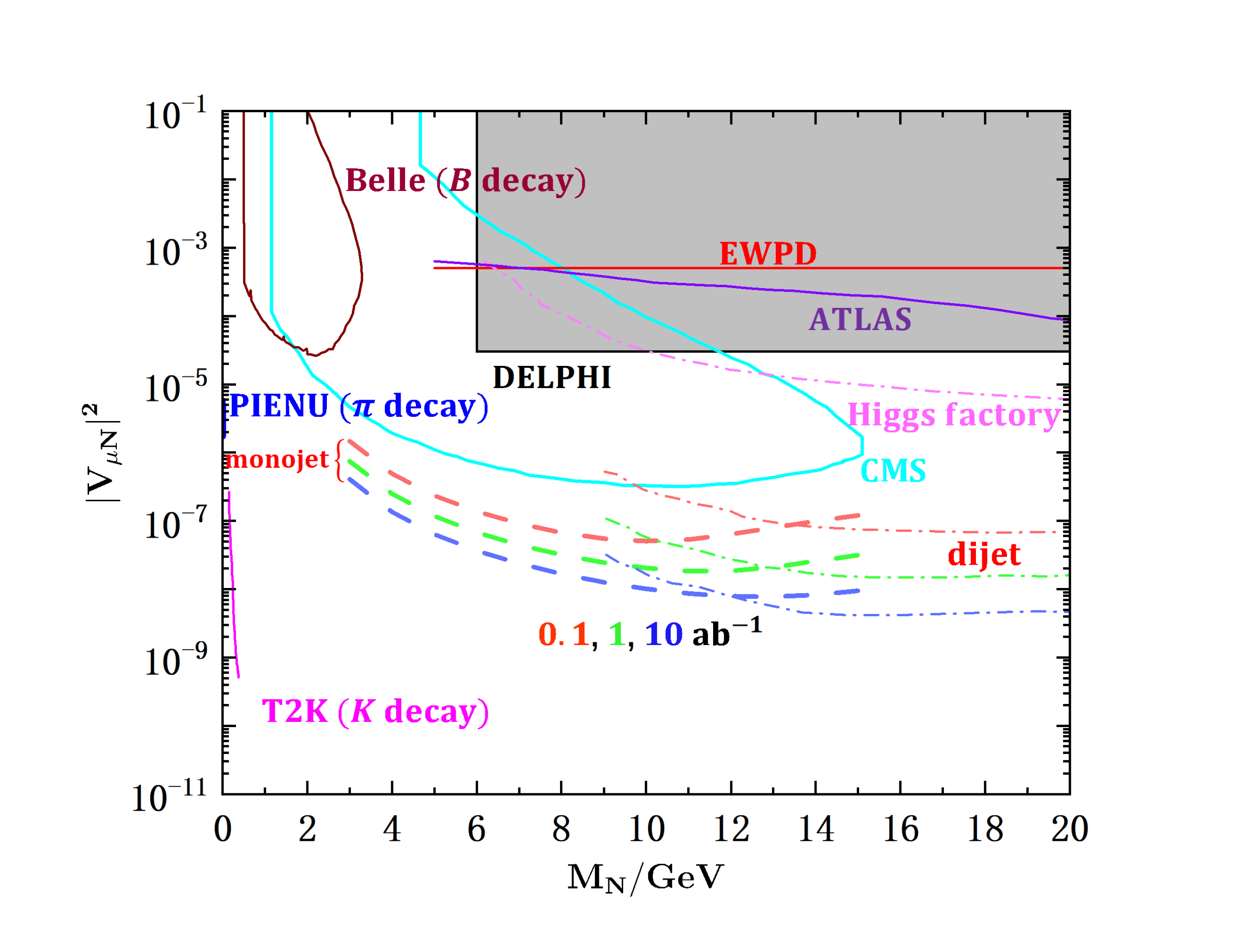}
\label{VmuN}
}
\caption{The upper bounds on $\left|V_{\ell N}\right|^{2}$ at 95\% CL from future $Z$-factories with three setups of integrated luminosities $L=0.1,1\ \&\ 10\ \text{ab}^{-1}$. Constraints from other experiments are also displayed for comparison, including direct searches by ALEPH~\cite{ALEPH:1991qhf}, DELPHI~\cite{DELPHI:1996qcc}, CMS~\cite{CMS:2022fut} and ATLAS~\cite{ATLAS:2019kpx}, indirect searches like $0\nu\beta\beta$~\cite{Deppisch:2020ztt}, electroweak precision data (EWPD)~\cite{Das:2018hph}, and meson decays measured by PIENU ($\pi$ decay)~\cite{PIENU:2019usb,PIENU:2017wbj}, T2K ($K$ decay)~\cite{T2K:2019jwa} and Belle  ($B$ decay)~\cite{Belle:2013ytx}, and expected constraints from future facilities such as dijet method at future $Z$-factories~\cite{Ding:2019tqq}, LNV processes searches at future Higgs factories ~\cite{Zhang:2018rtr}. The current experimental bounds are shown by solid curves, while the expected future bounds are shown by dashed curves.}
\label{VlN}
\end{figure}

In the assumption that the object heavy neutrino only participates in the interactions formulated in \eqref{eq:l} and only one mixing parameter $V_{\ell N}$ ($\ell=e$ or $\mu$) is nonzero, the cross section $\sigma(e^{-}e^{+}\rightarrow Z \rightarrow \nu N \rightarrow \nu \ell q^\prime \bar q)$ only depends on one unknown parameter $V_{\ell N}$ with the neutrino mass given. Therefore, the above upper bounds on the cross sections can be translated to the corresponding bounds on the mixing parameters $\left|V_{e N}\right|^{2}$ or $\left|V_{\mu N}\right|^{2}$. 
It was found that such neutrinos have very long lifetimes and large Lorentz boosts, and their typical decaying distances are comparable to detector sizes. If they decay after flying out of detectors, they will be undetectable. Therefore, the finite size of a detector must be taken into account. According to~\cite{Gronau:1984ct}, the decay probability of a heavy neutrino inside a detector reads
\begin{equation}
P=\mathrm{Exp}[\frac{l\cdot M_N}{p_N \cdot \tau_N }],
\end{equation}  
where $p_N$ is the 3-momentum of the heavy neutrino in the laboratory frame, $\tau_N$ is its lifetime given by the MadWidth~\cite{Alwall:2014bza} which is consistent with~\cite{Atre:2009rg}, and $l$ is taken to be the length from the interaction point to the calorimeter of the detector where the involved interactions can be almost fully reconstructed. In this work, we take the CEPC designation \cite{CEPCStudyGroup:2018rmc} as an example and set $l=1.6\,\mathrm{m}$ accordingly. Modifying the number of signal events $N_s$ in equation (4) by multiplying the probability factor $P$, we finally obtain the bounds on the maxing parameters $\left|V_{\ell N}\right|^2$ at 95\% CL as shown in Fig.~\ref{VlN}.

The current experimental constraints are also displayed, including direct searches in colliders~\cite{DELPHI:1996qcc,CMS:2022fut,ATLAS:2019kpx} and meson decays~\cite{PIENU:2019usb,PIENU:2017wbj,T2K:2019jwa,Belle:2013ytx}, and indirect searches~\cite{Das:2018hph,Deppisch:2020ztt}. Expected constraints to heavier neutrinos from future Higgs factories and $Z$-factories are also shown~\cite{Zhang:2018rtr,Ding:2019tqq}.

One can observe that the left halves of the monojet coupling curves rise more sharply than the corresponding cross section curves in FIG.~\ref{sigma}, which is reasonable because lighter neutrinos are more likely to decay outside of detectors. Despite this, the constraints on $|V_{\ell N}|^2$ for 3 GeV $<M_N<$ 15 GeV are still expected to be improved by at least one order of magnitude compared to the most stringent existing constraints given by direct searches at the DELPHI~\cite{DELPHI:1996qcc} and LHC~\cite{CMS:2022fut}, even with a conservative integrated luminosity. For the electron case, it seems that the current upper limit on $|V_{e N}|^2$ from $0\nu\beta\beta$ decay experiments is already as good. However, it should be noted that the model dependence of the indirect $0\nu\beta\beta$ constraint can make it invalid in some cases. For example, if there are other dynamics than heavy neutrinos contributing to $0\nu\beta\beta$ decays, then those different processes can interfere with each other, and the final restriction can be weakened considerably according to~\cite{Pascoli:2013fiz}. One can also see that our results can improve the constraints by at least one to two orders even compared with the recent CMS search for long-lived heavy neutrinos~\cite{CMS:2022fut} using the displaced-vertex information. In fact, it has been found that the displaced-vertex information at future $Z$-factories~\cite{Antusch:2017pkq} can make the constraints even stricter than our results, but again it depends on a strong assumption that the object heavy neutrino must have a sufficiently long lifetime to entirely escape from the SM background. This condition is not always satisfied. For instance, if the heavy neutrino can decay into a Majoron, its lifetime can be very short even with a tiny mixing parameter~\cite{Chikashige:1980ui}. In contrast, our results are only based on the proportional formula mentioned at the end of Sec.~II and hence more inclusive and general to other types of models or interactions. Besides, the background-free assumption in~\cite{Antusch:2017pkq} might be too optimistic.
From FIG.~\ref{VlN}, we can conclude that the monojet method for heavy neutrino searches is expected to be very efficient at future $Z$-factories, which will be able to fill the gap between the $|V_{\ell N}|^2$ constraints from meson decays in the small mass range and from hadron colliders in the large mass range.

Though we mainly focus on the situation that the heavy neutrino mix with the electron or muon sector, our framework can also be applied to the tau case with $N\rightarrow\tau q^\prime \bar q$, if we consider only the leptonic decay channel of tau $\tau \rightarrow \ell \nu \bar{\nu}$. Such a signal channel also produces a final state containing a monojet with a charged lepton inside and missing energy. Hence, the monojet method can also be applied, except that some selection conditions must be adjusted, such as the restriction on the reconstructed center-of-mass energy. With a similar analysis, we estimate that the constraint of the mixing parameter $|V_{\tau N}|^2$ is basically one order of magnitude looser than that in the electron or muon case. A more systematic analysis is left to future studies.

At the end of this section, we discuss some possible controversial details about the substructure of the monojet. One might be worried about the misidentification of the charged lepton especially electron in the jet, which, for example, would make it to difficult to differentiate the high-energy QED showering from the signal. To estimate such an effect, we in practice build \emph{micro-jets} from final state hadrons with $\Delta R <0.1$ first, reclassifying charged leptons as micro-jets, and then restart jet clustering and apply the previous cuts. The new results are comparable to the original ones, {e.g.}, the $|V_{e N}|^2$ upper limits with the integrated luminosity $0.1$ ab$^{-1}$ change from $2.2\times 10^{-7}$ to $3.3\times 10^{-7}$ and from $1.1\times 10^{-7}$ to $1.9\times 10^{-7}$ for $M_N=5$ GeV and $M_N=15$ GeV, respectively. Considering the parameters and the corresponding selection conditions of the micro-jet method can be adjusted more finely, and the performance of a future detector could be much better than the estimation here, we maintain the our original results and allow possible deterioration at most by a factor of 2. 

\section{Conclusion}
We have investigated the capacity of a future $Z$-factory for hunting heavy neutrinos $N$ considering a simple extension of the SM with a type-I seesaw neutrino masses sector. Focusing on the heavy neutrino mass range $M_N\in[3,15]$ GeV, the large Lorentz boost of the heavy neutrino motivated us to apply the monojet method to reconstruct the heavy neutrino signal. It was found that the substructure of the monojet played a crucial role in separating signal events from the SM background. Based on the simulated signal and background events with the corresponding event selection conditions applied, we have presented the future $Z$-factory constraints on the cross sections of the heavy neutrino signal events and the mixing parameters $|V_{\ell N}|^2$ with three setups of integrated luminosities $L=0.1,1,10$ ab$^{-1}$. It was observed that the sensitivity to the mixing parameters could reach approximately $\mathcal{O}(10^{-7})$ even with the most conservative integrated luminosity, improving the current direct experimental searches by one to two orders of magnitude. Therefore, we conclude that the monojet method in searching for 3-15 GeV heavy neutrinos at future $Z$-factories will be able to fill the gap left by searches in meson decays and hadron collider productions, which are more sensitive to lighter and heavier neutrinos, respectively. 

\section*{Acknowledgements}

The authors are grateful to Yichen Li and Manqi Ruan for inspiring discussions. 
This work is supported by Natural Science Foundation of China under grant No. 12005068 and 11975112.

\section*{Data Availability Statement}

This manuscript has no associated data or the data will not be deposited. [Authors’ comment: The number and sizes of the simulation events are too large to be uploaded. However, we’ve mentioned the necessary details of our simulation processes in Section III, which will be useful for future works.]

\end{document}